\documentclass[a4paper,11pt]{article}
\usepackage[utf8]{inputenc}

\usepackage{amsmath,amssymb,graphicx,epsfig,slashed}
\usepackage{color}

\usepackage[hdivide={30mm,,30mm}, vdivide={30mm,,30mm}, nohead]{geometry}

\newcommand{\De}{{\Delta}}
\newcommand{\Lm}{{\Lambda}}
\newcommand{\Om}{{\Omega}}

\newcommand{\ga}{{\gamma}}

\newcommand{\ep}{{\epsilon}}

\newcommand{\lm}{{\lambda}}

\newcommand{\sig}{{\sigma}}

\newcommand{\bbX}{{\mathbb{X}}}

\newcommand{\cD}{{\mathcal{D}}}
\newcommand{\cR}{{\mathcal{R}}}

\newcommand{\Slash}[1]{{\ooalign{\hfil/\hfil\crcr$#1$}}} 
\newcommand{\nn}{{\nonumber}}

\title{The cosmological constant in Supergravity}

\begin{document}

\thispagestyle{empty}

\begin{center}
{\Large{\textbf{The cosmological constant in Supergravity}}}
\\
\medskip
\vspace{1cm}
\textbf{
I.~Antoniadis$^{\,a,b,}$\footnote{antoniadis@itp.unibe.ch}, 
A.~Chatrabhuti$^{c,}$\footnote{dma3ac2@gmail.com}, 
H.~Isono$^{c,}$\footnote{hiroshi.isono81@gmail.com}, 
R.~Knoops$^{c,}$\footnote{rob.k@chula.ac.th}
}
\bigskip

$^a$ {\small Laboratoire de Physique Th\'eorique et Hautes \'Energies - LPTHE,\\ Sorbonne Universit\'e, CNRS, 4 Place Jussieu, 75005 Paris, France}

$^b$ {\small Albert Einstein Center, Institute for Theoretical Physics,
University of Bern,\\ Sidlerstrasse 5, CH-3012 Bern, Switzerland }

$^c$ {\small Department of Physics, Faculty of Science, Chulalongkorn University,
\\Phayathai Road, Pathumwan, Bangkok 10330, Thailand }

\end{center}

\vspace{1cm}

\begin{abstract}
 We propose a supersymmetrisation of the cosmological constant in ordinary $N=1$ supergravity that breaks supersymmetry spontaneously by a constant Fayet-Iliopoulos (FI) term associated to a $U(1)$ symmetry. This term is a variation of a new gauge invariant FI term proposed recently, which is invariant under K\"ahler transformations and can be written even for a gauged R-symmetry on top of the standard FI contribution. The two terms are the same in the absence of matter but differ in its presence. 
The proposed term is reduced to a constant FI-term up to fermion interactions that disappear in the unitary gauge in the absence of any F-term supersymmetry breaking. 
The constant FI term leads to a positive cosmological constant, uplifting the vacuum energy from the usual anti-de Sitter supergravity to any higher value.
\end{abstract}

\section{Introduction}

It is well known that the cosmological constant $\Lambda$ in supergravity is highly constrained. For given gravitino mass term $m_{3/2}$, there is a lowest value of $\Lambda=-3m_{3/2}^2$ corresponding to the anti de Sitter (AdS) supergravity, describing a massless spin-3/2 spinor in AdS~\cite{Freedman:2012zz}. It is obtained in the absence of matter fields by a constant superpotential. Uplifting this value breaks supersymmetry and can be done in principle dynamically by minimising a scalar potential. Supersymmetry breaking then occurs by a vacuum expectation value (VEV) of an F-auxiliary component of a chiral superfield containing the goldstino. 

In the absence of matter, one could still break supersymmetry by a VEV of a D-auxiliary component of a vector superfield which requires the addition of a constant Fayet-Iliopoulos (FI) contribution~\cite{Fayet:1974jb}. This can be done only when the vector superfield gauges the R-symmetry, under which the chiral compensator of $N=1$ supergravity becomes charged~\cite{Freedman:1976uk,Barbieri:1982ac}. A constant superpotential is however forbidden in that case, since it must be charged under R-symmetry, and there is no explicit gravitino mass term, although supersymmetry is broken in de Sitter (dS) space. In the presence of matter, a charged superpotenial can be written but then supesymmetry is also broken by the VEV of a chiral multiplet upon minimisation of the corresponding scalar potential.
Thus, the cosmological constant cannot be added as an independent parameter in supergravity for arbitrary breaking scale (gravitino mass).

An exception to the above situation is when supersymmetry is non-linearly realized by introducing a constrained  goldstino superfield $X$ satisfying the nilpotent condition $X^2=0$~\cite{Casalbuoni:1988xh,Komargodski:2009rz,Antoniadis:2014oya}. 
This eliminates the scalar component (sgoldstino) in terms of the goldstino bilinear and the scalar potential (in the absence of matter) becomes an arbitrary constant uplifting the minimal value 
$\Lambda=-3m_{3/2}^2$~\cite{Antoniadis:2014oya}. However in this case supersymmetry is not spontaneously broken in a linear way and the number of bosonic and fermionic degrees of freedom are not equal, invalidating in particular the usual ultraviolet properties of $N=1$ supergravity and of its low energy softly broken supersymmetric theory.

Recently, a new FI term was proposed that allows an arbitrary uplifting of the vacuum energy in the absence of matter fields~\cite{Cribiori:2017laj,Antoniadis:2018cpq}.\footnote{Another approach to the new FI term was proposed in \cite{Kuzenko:2018jlz}.} It does not require gauging the R-symmetry and in the unitary (super)gauge of massive gravitino, it is reduced to just an additive positive constant to $\Lambda$. In the presence of matter, however, it leads to an additional field-dependent contribution to the scalar potential. Moreover, it breaks the invariance of the standard two-derivative supergravity action under K\"ahler transformations.

In this work, we propose a modification/generalisation of this FI term that has the following properties: (1) it can be written independently whether the corresponding $U(1)$ gauges or not the R-symmetry; (2) in the absence of matter fields and in the case of an ordinary (non-R) $U(1)$, it coincides with the one proposed in~\cite{Cribiori:2017laj}; (3) in the case of a $U(1)_R$, it can be written on top of the standard constant FI term; (4) in the presence of matter fields the action is invariant under K{\"a}hler transformations and its bosonic contribution is always a constant FI term that uplifts in particular the vacuum energy by an arbitrary positive constant. 

The outline of the paper is the following. In Section~2, we review the new FI term and its properties. In Section~3, we present a modification/generalisation that is invariant under K\"ahler transformations. In Section~4, we compute its bosonic contribution to the standard supergravity action. Section~5 contains some concluding remarks. Finally, we have two appendices; Appendix A contains useful formulae used in the text, while in Appendix~\ref{Appendix:Fermions} we compute the fermionic part of the supergravity action.

\section{Review and definitions}

In~\cite{Cribiori:2017laj} a new FI term has been proposed of the form
\begin{align}
 \mathcal L_{\rm FI} &= -
 \xi \left[  S_0 \bar S_0 \frac{w^2 \bar w^2}{ T(\bar w^2) \bar T (w^2) } (V)_D \right]_D , \label{LFI}
 \end{align}
where we put the Planck mass to $1$, and $\xi$ is a constant parameter. We use the conventions of~\cite{Freedman:2012zz,Ferrara:2016een}.
Here $S_0 = \left( s_0, P_L \Omega_0, F_0 \right)$ and $\bar S_0 = \left( \bar s_0, P_R \Omega_0 , \bar F_0 \right)$ are the chiral (and anti-chiral) compensator fields with $(\text{Weyl}, \text{Chiral})$ weights $(1,1)$ and $(1,-1)$ respectively, and
$V$ is a real (vector) supermultiplet with weights $(0,0)$ and components $V = \left( v , \zeta, \mathcal H, A_\mu, \lambda, D \right)$,
where the first three components are zero in the Wess-Zumino gauge $v = \zeta = \mathcal H = 0$.
The linear projection $(V)_D$ has weights $(2,0)$ and is defined by 
\begin{align}
 (V)_D = ( D, ~ \slashed{\cal D} \lambda, ~ 0 , ~ {\cal D}^b \hat{F}_{ab}, ~ - \slashed{\cal D} \slashed{\cal D} \lambda, ~ - \Box^C {D}   \label{V_D}
 ) . 
\end{align}
The chiral (and anti-chiral) multiplets $w^2$ (and $\bar w^2$) are given by
\begin{align}
 w^2 = \frac{ \bar \lambda P_L \lambda}{S_0^2}, \ \ \ \ \ \ 
 \bar w^2 = \frac{  \lambda P_R \bar \lambda}{\bar S_0^2}, \label{w2def}
\end{align}
where the components of $\bar \lambda P_L \lambda$ are\footnote{Note that in this notation the field strength superfield $\mathcal W_\alpha$ is given by $\mathcal W ^2 = \bar \lambda P_L \lambda$, 
and $(V)_D$ corresponds to $\cal D^\alpha \mathcal W_\alpha$.} 
\begin{align}
\bar \lambda P_L \lambda = \Big( \bar \lambda P_L \lambda \  ;  \sqrt{2} P_L \big( - \frac{1}{2} \gamma \cdot  \hat F + i D \big) \lambda \  ; \ 
2 \bar \lambda P_L \slashed{\cal D} \lambda + \hat F^- \cdot \hat F^- - D^2 \Big) , 
\end{align}
where the self-dual and anti-self-dual  tensors are defined by
\begin{align}
\hat F_{\mu \nu}^\pm = \frac{1}{2} \left( \hat F_{\mu \nu} \pm \tilde{\hat F}_{\mu \nu} \right), \qquad
\tilde{\hat F}_{\mu\nu} = -\frac{1}{2}i \ep_{\mu\nu\rho\sig} \hat F^{\rho\sig},
\end{align}
and the definitions of the covariant field strength $\hat F$ and the covariant derivative $\slashed{\cal D} \lambda$ can be found in Appendix~\ref{Appendix:defs}. 
The conformal d'Alembertian is given by $\Box^C = \eta^{ab} {\cal D}_a {\cal D}_b $.
Note that $w^2$ has weights $(1,1)$, $\bar w^2$ has weights $(1,-1)$, and $\bar \lambda P_L \lambda$ has weights $(3,3)$. 

The chiral (and anti-chiral) projection operators $T$ (and $\bar T$) are defined in~\cite{Kugo:1983mv,Ferrara:2016een}. 
In particular if $\cal C$ is a general (unconstrained) multiplet of weights $(\omega, \omega - 2)$ given by
\begin{align}
 {\cal C} = \Big( {\cal C}, {\cal Z}, {\cal H}, {\cal K}, {\cal B} _a , \Lambda, {\cal D} \Big) ,
\end{align}
then $T (\cal C)$ has weights $(\omega + 1, \omega +1)$ and is given by
\begin{align}
T ({\cal C}) = \Big( - \frac{1}{2} {\cal K} , - \frac{1}{2} \sqrt{2} i P_L (\slashed{\cal D} {\cal Z} + \Lambda ) , \frac{1}{2} ( {\cal D} + \Box^C {\cal C} + i {\cal D}_a {\cal B}^a ) \Big) .
\label{Tdef}
\end{align}
The resulting chiral multiplet $T(C)$ has weights $(\omega + 1, \omega + 1)$.
The operation $T$ acting on a chiral multiplet $X = (\phi, P_L \chi, F)$ vanishes, i.e. $T(X) = 0$, 
while its action on an anti-chiral multiplet $\bar X = \left( \bar \phi, P_R \chi , \bar F \right)$ of weights $(1,-1)$ is defined as
\begin{align}
 T(\bar X) = \left( \bar F , \slashed{\cal D} P_R \chi , \Box^C \bar \phi \right). \label{Tchiral}
\end{align}
For more information, the reader is referred to Appendix~\ref{Appendix:defT}.
In rigid supersymmetry, this corresponds to the usual chiral (and anti-chiral) projection operators $\bar D^2$ (and $D^2$).

For simplicity, we assume a constant gauge kinetic function. The kinetic terms for the gauge multiplet are given by
\begin{align}
 \mathcal L_{\text{kin}} = - \frac{1}{4} \left[ \bar \lambda P_L \lambda \right]_F.  \label{Lkin}
\end{align}
The extension to a non-trivial gauge kinetic function is given in~\cite{Antoniadis:2018cpq}. 

The operation $[ \ \ ]_F$ acts on a chiral multiplet $X=(\phi,P_L\Omega,F)$ with weights $(3,3)$, giving~\cite{Ferrara:2016een}\footnote{
Note that the definitions of $[ \ \ ]_F$ and $[ \ \ ]_D$ below do not involve the spacetime integral $\int d^4x$, which appears in the corresponding expressions in~\cite{Ferrara:2016een}.}
\begin{align}
[X]_F = e \bigg[ F+\frac{1}{\sqrt{2}}\bar\psi_\mu \ga^\mu P_L\Omega + \frac{1}{2}\phi\bar\psi_\mu \ga^{\mu\nu}P_R\psi_\nu \bigg] + {\rm h.c.} \, .
\end{align}
Note that this already contains the Hermitian conjugate.
The operation $[ \ \ ]_D$ acts on a real multiplet 
$C = \left( C , \zeta, \mathcal H, v_\mu, \lambda, D \right)$
of weights $(2,0)$, giving~\cite{Ferrara:2016een}
\begin{align}
 [ C ]_D = \frac{1}{2}
 e \bigg[
 &D - \frac{i}{2} \bar \psi \cdot \gamma \gamma_* \lambda - \frac{1}{3} C R(\omega) 
 + \frac{1}{6} (C \bar \psi_\mu \gamma^{\mu \rho \sigma} - i \zeta \gamma^{\rho \sigma} \gamma_* ) R'_{\rho \sigma} (Q)  \notag \\
 &+ \frac{1}{4} \epsilon^{abcd} \bar \psi_a \gamma_b \psi_c \bigg(v_d - \frac{1}{2} \bar \psi_d \zeta \bigg) 
 \bigg].
 \label{[C]D}
\end{align}
Here $\psi$ is the gravitino, and $R(\omega)$ and $R'_{\rho \sigma} (Q)$ are the graviton and gravitino curvatures.

The Lagrangian contains a term
\begin{align}
\mathcal L_{\text{FI}} / e  = -\xi s_0 \bar s_0 D .
 \label{onlyFI_VD}
\end{align}
After the auxiliary field $D$ is integrated out with the contribution from the kinetic term $D^2/2$ taken into account, the scalar potential contains a term proportional to $\xi^2$,
\begin{align}
\mathcal V_{\text{FI}}  =  \frac{ \xi^2}{2}  \left( s_0 \bar s_0 \right)^2 .
\end{align}
In the absence of additional matter fields, one can use the Poincar\'e gauge $s_0 = \bar s_0 = 1$,
resulting in a constant D-term contribution to the scalar potential. 
However, when matter fields are included, the Einstein frame gauge gives $s_0 = \bar s_0 = e^{K/6}$  fixing the conformal symmetry, leading to a field dependent FI contribution to the scalar potential:
\begin{align}
 \mathcal V_{\text{FI}} =  \frac{ \xi^2}{2}  e^{2K/3} . \label{onlyFI_VD2} 
\end{align}
The implications of such a term to inflation have been studied in~\cite{Aldabergenov:2017hvp,Antoniadis:2018cpq}.

In the presence of matter fields the FI term (\ref{LFI}) is not invariant under K\"ahler transformations. 
The purpose of this paper is to construct a term, similar to eq.~(\ref{LFI}), invariant under K\"ahler transformations. 
As a consequence, the contribution to the scalar potential will no longer depend on $e^{2K/3}$ as in eq.~(\ref{onlyFI_VD2}).

\section{K\"ahler invariant generalization of the new FI term}

In this section we find a generalization of the new FI term in eq.~(\ref{LFI}) that is invariant under K\"ahler transformations in the presence of matter multiplets. 
For simplicitly, we denote generically the chiral multiplets by $X$. The standard $N=1$ supergravity Lagrangian is given by
\begin{align}
 \mathcal L_X &= -3 \left[ S_0 \bar S_0 e^{- K/3} \right]_D + \left[ S_0^3 W \right]_F , 
 \label{LX}
\end{align}
for a K\"ahler potential $K(X, \bar X)$ and a superpotential $W(X)$. A K\"ahler transformation with parameter $J(X)$ is given by
\begin{align}
 K(X , \bar X) &\rightarrow K(X ,\bar X)  + J(X) + \bar J(\bar X), \notag \\
 W(X) &\rightarrow W(X) e^{-J(X)} , \notag \\
  S_0 &\rightarrow  S_0 e^{\frac{J(X)}{3}} \label{chiralKahler} . 
\end{align}
It is clear that $\mathcal{L}_X$ in eq.~(\ref{LX}) is invariant under K\"ahler transformations. However,  
the new FI term proposed in~\cite{Cribiori:2017laj} and given in eq.~(\ref{LFI}) is not. 

In~\cite{Cribiori:2017laj} it is suggested that a K\"ahler invariant generalization can be found by making the FI constant field dependent, i.e. $\xi = \xi (X, \bar X)$. 
However, $\bar T (w^2)$ is not K\"ahler covariant since the conformal compensator $S_0$ transforms under K\"ahler transformations. 
Thus, under a K\"ahler transformation, $w^2$ transforms as $w^2 \rightarrow w^2 e^{-2J/3}$ while $\bar T ( w^2 e^{-2J/3} ) \neq \bar T (w^2) e^{-2J/3} $. 
Instead, in this paper we keep $\xi$ constant, 
and modify the new FI term by requiring the compensator fields $S_0$ and $\bar S_0$ to appear in the new FI term only through the K\"ahler invariant combination $S_0 \bar S_0 e^{-K/3}$.

First, recall that the operators $T$ and $\bar T$ should act on multiplets with weights $(\omega, \omega-2)$ and $(\omega, 2-\omega)$ respectively.
 We therefore want to replace $\bar w^2$ with a multiplet proportional to $\lambda P_R \bar \lambda$ with weights $(\omega, \omega-2)$. 
 Since $\lambda P_R \bar \lambda$ is K\"ahler invariant and has weights $(3,-3)$, we should multiply it with the K\"ahler invariant 
 combination $S_0 \bar S_0 e^{-K/3}$ of weights $(2,0)$ to obtain a multiplet of weights $(\omega , \omega-2)$.
 Indeed
 \begin{align} 
 \bar  w'^2 = (S_0 \bar S_0 e^{-K/3})^m \lambda P_R \bar \lambda
 \end{align}
has weights $(2m +3, -3)$. 
Thus, $(2m + 3, -3 ) = (\omega , \omega-2)$ can be solved for $\omega= -1$ and $ m = -2 $, and
one can define a K\"ahler invariant combination $\bar w'^2$ with weights $(-1,-3)$,
\begin{align}
 \bar w'^2 = \frac{ \bar \lambda P_R \lambda}{(S_0 \bar S_0 e^{-K/3})^2} . \label{wprimedefbar}
\end{align}
The resulting $T(\bar w'^2)$ has weights $(0,0)$. Similarly, one can construct $\bar T (w'^2)$ with weights $(0,0)$,
where
\begin{align}
 w'^2 = \frac{ \bar \lambda P_L \lambda}{(S_0 \bar S_0 e^{-K/3})^2} . \label{wprimedef}
\end{align}

By the same arguments, the new FI contribution to the Lagrangian has the form
\begin{align}
 \mathcal L_{\rm FI,\text{new}} = -
 \xi  \left[ (S_0 \bar S_0 e^{-K/3} )^k \frac{(\bar \lambda P_L \lambda)(\lambda P_R \bar \lambda)}{T(\bar w'^2) \bar T(w'^2) } (V)_D \right]_D . \label{LFInew}
\end{align}
The operation $[ \ \ ]_D$ is defined only on a multiplet with weights $(2,0)$, 
from which it follows that $k=-3$. 

We conclude that the new FI term is given by 
\begin{align}
 \mathcal L_{\rm FI,\text{new}} = -
 \xi \left[ (S_0 \bar S_0 e^{-K/3} )^{-3} \frac{(\bar \lambda P_L \lambda)( \lambda P_R \bar \lambda)}{T(\bar w'^2) \bar T(w'^2) } (V)_D \right]_D , \label{LFIp}
\end{align}
with $w'^2$ and $\bar w'^2$ defined in eqs.~(\ref{wprimedefbar}) and (\ref{wprimedef})
that are invariant under K\"ahler transformations and have the correct Weyl and Chiral weights. 
It remains to be shown in the next section that this indeed leads to a constant D-term contribution to scalar potential.

\section{A constant FI contribution to the scalar potential}

In this section we calculate the (purely) bosonic contributions to the D-term scalar potential of the Lagrangian
\begin{align}
 \mathcal L = \mathcal L_X + \mathcal L_{\text{kin}} + \mathcal L_{\rm FI,\text{new}},
\end{align}
with the matter contributions $\mathcal L_X$ given by eq.~(\ref{LX}), 
the gauge kinetic terms $\mathcal L_{\text{kin}}$ given by eq.~(\ref{Lkin}), and the new FI term $\mathcal L_{\rm FI, \text{new}}$ defined in eq.~(\ref{LFIp}).

For compactness, we put the fermions to zero, and postpone the treatment of the fermion couplings to Appendix~\ref{Appendix:Fermions}.
The purely bosonic contributions to $\bar \lambda P_L \lambda$ are
\begin{align}
 \bar \lambda P_L \lambda = \Big( 0, ~  0, ~  F^- \!\cdot\! F^- - D^2  \Big),
\end{align}
while the purely bosonic contributions to $w'^2$ are given by\footnote{Note that $w'^2$ has seven components since $w'^2$ is neither real nor chiral. As for $\bar w'^2$, its third component vanishes and instead the fourth one becomes the complex conjugate of the third one of $w'^2$.
}
\begin{align}
 w'^2 &= \Bigg(0, ~ 0, ~
 2\big( s_0 \bar s_0 e^{-K/3} \big)^{-2}(D^2 - F^- \!\cdot\! F^-),
 ~ 0, \nn\\
& \qquad \qquad ~ 0, ~ 0,
  ~ 4\big( s_0 \bar s_0 e^{-K/3} \big)^{-2} \left(D^2 - F^- \!\cdot\! F^-  \right) 
  \bigg(  \frac{\bar F_0}{\bar s_0} - \frac{K_{\bar\phi}\bar F}{3} \bigg)\Bigg).
\end{align}
The bosonic contributions to the anti-chiral projection $\bar T(w'^2)$ are
\begin{align}
 \bar T ( w'^2) = \Bigg( & 
 \big(  s_0 \bar s_0 e^{-K/3} \big)^{-2} \left( F^- \!\cdot\! F^- - D^2  \right),  ~
 0, ~ \notag \\
& \quad 
2\big( s_0 \bar s_0 e^{-K/3} \big)^{-2} \left(D^2 - F^- \!\cdot\! F^-  \right) 
\bigg(  \frac{\bar F_0}{\bar s_0} - \frac{K_{\bar\phi}\bar F}{3} \bigg) \Bigg) . \label{Tbar}
\end{align}
Here the chiral multiplet $X$ is defined as $X = (\phi , P_L\chi, F)$ and $K_\phi = \partial_\phi K$.

Next, we notice that the real multiplet
\begin{align}
 \mathcal R = (S_0 \bar S_0 e^{-K/3} )^{-3} \frac{(\bar \lambda P_L \lambda)(\bar \lambda P_R \lambda)}{T(\bar w'^2) \bar T(w'^2) }
\end{align}
is a function of chiral multiplets $Z^\alpha = X, S_0, \bar \lambda P_L \lambda, T(\bar w'^2)$ and their anti-chiral counterparts $\bar Z^{\bar\alpha}$. 
Its components are
\begin{align}
\label{R-comp}
 \mathcal R = \Big(
 & \mathcal R,  ~ 0, 
 ~ -2 \mathcal R_\alpha F^\alpha, 
 ~ i\mathcal R_\alpha {\cal D}_\mu Z^\alpha - i \mathcal R_{\bar \alpha} {\cal D}_\mu \bar Z^{\bar \alpha},  
 ~ 0, \ \notag \\
  & ~ 2 \mathcal R_{\alpha \bar\beta} \big( - {\cal D}_\mu Z^\alpha {\cal D}^\mu \bar Z^{\bar \beta} + F^\alpha \bar F^{\bar \beta}  \big)  
 \Big) ,
\end{align}
where $\mathcal R_\alpha = \frac{\partial \mathcal R}{\partial Z^\alpha}$ and all fields are replaced by their lowest components. Note also that obviously fermionic contributions are ignored in eq.~\eqref{R-comp}.
As a result, the components of $\mathcal R$ only with bosonic fields are given by
\begin{align}
 \mathcal R = \Big( 0, ~ 0, ~ 0, ~ 0, ~  0,  ~ 2 s_0 \bar s_0 e^{-K/3}  \Big) .
\end{align}
It follows that the contribution to the new FI term Lagrangian eq.~(\ref{LFIp}) is given by 
\begin{align}
 \mathcal L_{\rm FI,\text{new}}/e = - \xi s_0 \bar s_0 e^{-K/3} D .
\end{align}
In the Einstein frame gauge $s_0 = \bar s_0 = e^{K/6}$, this becomes a constant FI term
\begin{align}
 \mathcal L_{\rm FI,\text{new}}/e = -\xi D .
\end{align}
We therefore conclude that the Lagrangian of the ${\rm U}(1)$ gauge field sector is
\begin{align}
 \left( \mathcal L_{\text{kin}} + \mathcal L_{\rm FI,\text{new}} \right) /e  = -\frac{1}{4} F_{\mu \nu} F^{\mu \nu} + \frac{1}{2} D^2 - \xi D + \text{fermions} , 
\end{align}
which results in a constant FI contribution to the scalar potential. However, the terms in the denominator of eq. (\ref{LFIp}) are proportional to $D^2 - F^- \cdot F^-$ and $D^2 - F^+ \cdot F^+$ (see eq.~(\ref{Tbar})). The new FI term is therefore local only if $\langle D \rangle$ is non-vanishing.
The theory becomes ill-defined as $ \xi \rightarrow 0$, as was the case in~\cite{Cribiori:2017laj},
since in this limit $\langle D \rangle=0$.

However, in contrast with the term proposed by~\cite{Cribiori:2017laj}, the term proposed in eq.~(\ref{LFIp}) is manifestly K\"ahler and Weyl invariant. While both terms can be easily extended to include charged matter fields,
only eq.~(\ref{LFIp}) is consistent with matter fields that are charged under a gauged R-symmetry, as a consequence of its K\"ahler invariance. 

Therefore, while in~\cite{Antoniadis:2018cpq}, the new FI term of~\cite{Cribiori:2017laj} could only be added on top of the the usual FI contribution in the K\"ahler frame where the gauge symmetry is not an R-symmetry, the new FI term in eq.~(\ref{LFIp}) is consistent with the usual FI contribution in any K\"ahler frame, resulting in two constant contributions to the D-term contribution in the scalar potential.

A few remarks are in order:
\begin{itemize}
\item
Firstly, notice that one could also obtain a constant FI contribution to the D-term by making the substitution
$S_0 \rightarrow S_0 e^{-K/6}$ and $\bar S_0 \rightarrow \bar S_0 e^{-K/6}$ in eq.~(\ref{LFI}), giving
\begin{align}
 \mathcal L_{\rm FI,\text{c}} = -
 \xi \left[ S_0 \bar S_0 e^{-K/3} \frac{w_c^2 \bar w_c^2}{T(\bar w_c^2) \bar T(w_c^2) } (V)_D  \right]_D,
\end{align}
with
\begin{align}
 w_c^2 = \frac{\bar \lambda P_L \lambda }{S_0^2 e^{-K/3} }.
\end{align}
The resulting Lagrangian indeed contains the term $ \xi S_0 \bar S_0 e^{-K/3} D$ which results in a constant FI term in the Einstein frame gauge.
However, this term is not invariant under K\"ahler transformations eqs~(\ref{chiralKahler}) since $T(\bar{w}_c^2)$ does not transform covariantly, and we therefore do not analyse this term further.

\item Secondly, note that in the absence of matter fields (and therefore $K(X , \bar X) = 0$), $w^2$ defined in eq.~(\ref{w2def}) and $w'^2$ defined in eq.~(\ref{wprimedef}) are related by
\begin{align}
 w'^2 = \bar S_0^{-2} w^2 .
\end{align}
By using the property of the anti-chiral projection operators that for an anti-chiral field $\bar Z$ and a multiplet ${\cal C}$,
\begin{align}
 \bar T ( \bar Z {\cal C} ) = \bar Z \bar T ( {\cal C}),
\end{align}
we find that 
\begin{align}
 \bar T (w'^2) =  \bar S_0^{-2} \bar T (w^2),
\end{align}
and similarly
\begin{align}
 T (\bar w'^2) = S_0^{-2} T (\bar w^2) .
\end{align}
As a result, in the absence of matter fields, our proposed FI term in eq.~(\ref{LFInew}) is identical to the one proposed in~\cite{Cribiori:2017laj} 
and given in eq.~(\ref{LFI}).\footnote{We thank A.~Van Proeyen for bringing our attention to this.}
\end{itemize}

\section{Conclusions}

In summary, in this work, we proposed a supersymmetrisation of the cosmological constant in $N=1$ supergravity, arising from a constant FI term associated to an abelian gauge symmetry as in global supersymmetry. 
In contrast to the standard FI term which requires the gauging of R-symmetry, it can be written for any $U(1)$. 
It is obtained by a variation of a new FI term proposed recently in a way that is invariant under K\"ahler transformations, leading to just a constant FI term up to fermion contributions that disappear in the unitary gauge in the absence of any F-term supersymmetry breaking. 

Since a generalisation of such `new' FI terms is not unique and may in general involve new field dependent functions\footnote{
It is worth noting that shortly after submitting this paper, several attempts in this direction have been made. See for example \cite{Farakos:2018sgq} and \cite{Aldabergenov:2018nzd}.
}, an interesting question is whether they can arise in the effective supergravity of string compactifications and what their form is. An obvious application of the proposed term is that it uplifts the vacuum energy with a positive contribution, allowing to realise `realistic' models of moduli stabilisation and inflation based on the KKLT mechanism~\cite{KKLT} without the need of introducing anti-D3 branes, using a $U(1)$ factor of `effective' 3-branes gauge group whose gauge coupling is fixed by the ten-dimensional dilaton~\cite{Antoniadis:2006eu,Antoniadis:2008uk}.

\section*{Acknowledgements}

We would like to thank Jean-Pierre Derendinger, Sergey Ketov, J. Anibal Sierra-Garcia and Antoine Van Proeyen for fruitful discussions.
This work was supported in part by the Swiss National Science Foundation, in part by a CNRS PICS grant and in part by the ``CUniverse'' research promotion project by Chulalongkorn University (grant reference CUAASC). 

\appendix

\section{Useful formulas} \label{Appendix:defs}

\subsection{The chiral projection and complex multiplets} \label{Appendix:defT}

This Appendix is based on~\cite{Ferrara:2016een}.
The operation $T$ acts on a complex multiplet $\cal C$ with weights $(\rm{Weyl,Chiral}) = (\omega, \omega - 2)$, producing a chiral multiplet with the first component $-\frac{1}{2} \cal K$.
In particular if $\cal C$ is a general (complex) multiplet given by
\begin{align}
 {\cal C} = \Big( {\cal C}, {\cal Z}, {\cal H}, {\cal K}, {\cal B} _a , \Lambda, {\cal D} \Big) ,
\end{align}
then $T (\cal C)$ has weights $(\omega + 1, \omega +1)$ and is given in eq.~(\ref{Tdef}), repeated here for convenience\footnote{ 
The three-component notation of chiral/anti-chiral multiplets will be defined shortly.}
\begin{align}
T ({\cal C}) = \Big( - \frac{1}{2} {\cal K} , - \frac{1}{2} \sqrt{2} i P_L (\slashed{\cal D} {\cal Z} + \Lambda ) , \frac{1}{2} ( {\cal D} + \Box^C {\cal C} + i {\cal D}_a {\cal B}^a ) \Big) .
\end{align}
The anti-chiral projector $\bar T ({\cal C})$ acts on a multiplet of weights $(\omega, 2 - \omega)$ and results in an anti-chiral multiplet of weights $(\omega + 1, - \omega - 1)$ with lowest component $ - \frac{1}{2} {\cal H} $.

The restriction of ${\cal C} = C$ is real produces a real multiplet. This also implies that the Chiral weight $c=0$. Moreover, ${\cal Z} = \eta$ and $\Lambda = \lambda$ are Majorana $(P_R {\cal Z} )^C = P_L {\cal Z} $,
and ${\cal K} = \bar {\cal H}$, while ${\cal B}_\mu = B_\mu$ and ${\cal D} = D$ are real,
\begin{align}
 C = \Big( C, \zeta, {\cal H}, \bar {\cal H}, B_\mu, \lambda, D \Big).
\end{align}
For a real multiplet, we usually abbreviate $\bar {\cal H}$. The chiral projector $T$ can only act on a real multiplet of weights $(2,0)$.

A chiral multiplet is obtained by the restrictions
\begin{align}
 P_R {\cal Z} = 0, \ \ {\cal K} = 0, \ \ {\cal B}_\mu = i {\cal D}_\mu {\cal C}, \ \ \Lambda = 0, \ \ {\cal D} = 0 ,
\end{align}
and is given by
\begin{align}
 X = \Big( X, -i \sqrt{2} P_L \chi, -2 F, 0 , i {\cal D}_\mu X, 0 , 0  \Big). \label{chiralmultiplet}
\end{align}
Similarly an antichiral multiplet is given by
\begin{align}
 \bar X = \Big( \bar X, i \sqrt{2} P_R \chi, 0 , -2 \bar F, - i {\cal D}_\mu \bar X, 0 , 0  \Big) . \label{antichiralmultiplet}
\end{align}
However, the chiral multiplet in eq.~(\ref{chiralmultiplet}) is usually denoted as $X = (X, P_L \chi, F)$, while the anti-chiral multiplet $\bar X = (\bar X, P_R \chi, \bar F)$. 
A chiral multiplet has equal chiral and Weyl weights $c = \omega$, and an anti-chiral multiplet has $c = -\omega$.
As a result, $T$ can only act on an anti-chiral multiplet of weights $(1,-1)$ and is given by
\begin{align}
 T(\bar X) = \left( \bar F , \slashed{\cal D} P_R \chi , \Box^C \bar X \right). \label{TchiralAppendix}
\end{align}

\subsection{Covariant derivatives}

The definitions of $\slashed{\cal D} \lambda$ and the covariant field strength $\hat F_{ab}$ can be found in eq. (17.1) of \cite{Freedman:2012zz}, 
which reduce for an abelian gauge field to
\begin{align}
 \hat F_{ab} 
 &= e_a^{\ \mu} e_b^{\ \nu} \left( 2 \partial_{[\mu} A_{\nu ]} + \bar \psi_{[\mu} \gamma_{\nu ]} \lambda \right)
 = F_{ab} + e_a^{\ \mu} e_b^{\ \nu}\bar \psi_{[\mu} \gamma_{\nu ]}\lambda, \notag \\
 {\cal D}_\mu \lambda &= \left( \partial_\mu - \frac{3}{2} b_\mu + \frac{1}{4} w_\mu^{ab} \gamma_{ab} - \frac{3}{2} i \gamma_* \mathcal A_\mu \right) \lambda
 -\left( \frac{1}{4} \gamma^{ab} \hat F_{ab} + \frac{1}{2} i \gamma_* D \right) \psi_\mu .
\end{align}
Here, $e_a^{\ \mu}$ is the vierbein, with frame indices $a,b$ and coordinate indices $\mu, \nu$.
The fields $w_\mu^{ab}$, $b_\mu$, and $\mathcal A_\mu$ are the gauge fields corresponding to Lorentz transformations, dilatations, and $T_R$ symmetry of the conformal algebra respectively,
while $\psi_\mu$ is the gravitino. The covariant derivatives of other fields can be found in eq.~(4.6) of \cite{Ferrara:2016een}.

\subsection{Multiplication laws}

Given a set of (complex) multiplets ${\cal C}^i$, one can construct a new multiplet $\tilde {\cal C} = f({\cal C}^i)$ with components
\begin{align}
\tilde{ \cal C} &= f , \notag \\
\tilde{ \cal Z} &= f_i {\cal Z}^i , \notag \\
\tilde{ \cal H} &= f_i {\cal H}^i - \frac{1}{2} f_{ij} \bar {\cal Z}^i P_L {\cal Z}^j , \notag \\
\tilde{ \cal K} &= f_i {\cal K}^i - \frac{1}{2} f_{ij} \bar {\cal Z}^i P_R {\cal Z}^j , \notag \\
\tilde{ \cal B}_\mu &= f_i {\cal B}_\mu^i + \frac{1}{2} i f_{ij} \bar {\cal Z}^i P_L \gamma_\mu {\cal Z}^j , \notag \\
\tilde{ \Lambda} &= f_i \Lambda^i + \frac{1}{2} f_{ij} \left[ i \gamma_* \slashed{\cal B}^i + P_L {\cal K}^i + P_R {\cal H}^i - \slashed{\cal D} {\cal C}^i \right] {\cal Z}^j - \frac{1}{4} f_{ijk} {\cal Z}^i \bar {\cal Z}^j {\cal Z}^k , \notag \\
\tilde{ \cal D} &= f_i {\cal D}^i + \frac{1}{2} f_{ij} \left( {\cal K}^i {\cal H}^j - {\cal B}^i \cdot {\cal B}^j - {\cal D} {\cal C}^i \cdot {\cal D} {\cal C}^j - 2 \bar \Lambda^i {\cal Z}^j - \bar {\cal Z}^i \slashed{\cal D} {\cal Z}^j \right) \notag \\
& \ \ \ \ - \frac{1}{4} f_{ijk} \bar {\cal Z}^i \left( i \gamma_* \slashed{\cal B}^j + P_L {\cal K}^j + P_R {\cal H}^j \right) {\cal Z}^k + \frac{1}{8} f_{ijkl} \bar {\cal Z}^i P_L {\cal Z}^j \bar {\cal Z}^k P_R {\cal Z}^l .
\label{multiplicationlaw}
\end{align}
For the multiplication laws involving real and (anti-)chiral multiplets, see section 5 of~\cite{Ferrara:2016een}.
The bar on spinors are always the Majorana conjugate, $\bar\chi = \chi^T\hat{C}$, where $\hat{C}$ is the charge conjugation matrix satisfying $\hat C \ga_\mu \hat C^{-1}=-\ga_\mu^T$. We use this conjugate even if the spinor is not Majorana. This is the convention in~\cite{Ferrara:2016een}. 

\subsection{Components}
For convenience, we here summarise the multiplets in the seven-component notation.
The multiplet $\cR$ is made up of the following multiplets
\begin{align}
X &= (\phi, ~ -i\sqrt{2}P_L\chi, ~ -2F, ~ 0, ~ +i\cD_\mu\phi, ~ 0, ~ 0), \\
S_0 &= (s_0, ~ -i\sqrt{2}P_L\Om_0, ~ -2F_0, ~ 0, ~ +i\cD_\mu s_0, ~ 0, ~ 0), \\
\bar\lm P_L\lm &= (\bar\lm P_L\lm, ~ -i\sqrt{2}P_L\Lm, ~ 2D_-^2, ~ 0, ~ +i\cD_\mu(\bar\lm P_L\lm), ~ 0, ~ 0), \\
(V)_D &= (D, ~ \Slash{\cD}\lm, ~ 0, ~ 0, ~ \cD^b\hat F_{ab}, ~ -\Slash{\cD}\Slash{\cD}\lm, ~ -\square^CD),
\end{align}
and their charge conjugates.
The chiral multiplet $w'^2$ is defined by eq.~\eqref{w2def}.

\section{Fermion couplings} \label{Appendix:Fermions}

This appendix presents the fermionic terms in the new FI contribution \eqref{LFIp} up to the quadratic order in fermions.
Concerning the fermion couplings, the contribution to the Lagrangian from~(\ref{LFIp}) is given by \cite{Freedman:2012zz}
\begin{align}
 \mathcal L_{\rm{FI,new}} / e &= {-\frac{\xi}{2}}\left[ (\mathcal R)_{\cal D} D - {\cal D}^b \hat {F}_{ab} (\mathcal R)^a_{\cal B} - { ( \bar{\mathcal R} )_\lambda} \slashed{\cal D} \lambda 
 - \frac{1}{2} i D \psi \cdot \gamma \gamma_* (\mathcal R)_\lambda
 + \dots, \right],
\end{align}
which is obtained by applying eq.~\eqref{[C]D} to the definition $\mathcal L_{\rm{FI, new}} / e = {-\xi}[\mathcal R \!\cdot\! (V)_D ]_D$ with 
\begin{align}
 \mathcal R = (S_0 \bar S_0 e^{-K/3} )^{-3} \frac{(\bar \lambda P_L \lambda)(\bar \lambda P_R \lambda)}{T(\bar w'^2) \bar T(w'^2) } ,
\end{align}
and $(V)_D$ given by eq.~(\ref{V_D}).
We therefore need the ${\cal B}_\mu$, $\lambda$, and $D$ components of $\mathcal R$. 
Since we are interested in terms with two fermions in the new FI contribution, we need terms with up to two fermions for $(\cR)_{\mathcal{B}}, (\cR)_D$, and terms with one fermion for $(\cR)_\lm$. It therefore turns out that we need terms with up to two fermions for the lowest component of $T(\bar w'^2)$, terms with one fermion for the fermionic (i.e. second) component of $T(\bar w'^2)$, and terms with no fermions for the $F$-component of $T(\bar w'^2)$. 
Let us denote these component fields by
\begin{align}
T(\bar w'^2) = (C_T, ~ P_L\chi_T, ~ F_T).
\end{align}
Their explicit forms are given by
\begin{align}
\De^2 C_T &= -[ (D_{+}^2)_{\rm 0f} + \bar\bbX_{\rm 2f} ], \\
\De^2 P_L\chi_T &= P_L\bigg[ 
- \Slash\cD_{\rm 0f}\Lm_{\rm 1f} 
+ 2\bigg( \frac{\Slash\cD_{\rm 0f} s_0}{s_0}
- \frac{K_{\phi}\Slash\cD_{\rm 0f}\phi}{3} \bigg) \Lm_{\rm 1f}
- \frac{2K_{\phi}(D_+^2)_{\rm 0f}}{3} \chi 
\bigg], \\
\De^2 F_T &= 
\bigg( \frac{2}{s_0}F_0 - \frac{2K_{\phi}}{3} F_\phi \bigg) (D_{+}^2)_{\rm 0f},
\end{align}
where the subscript ${\rm 2f}$ indicates the two-fermion parts of the relevant term, and the same definition applies to the subscripts ${\rm 0f}, {\rm 1f}$. 
We also introduced the following symbols for compactness of the formulae,
\begin{align}
K_\phi &= \frac{\partial K(\phi, \bar\phi)}{\partial \phi}, \\
\De &= s_0\bar s_0 e^{-K/3}, \\
P_L\Lm &= \sqrt{2}P_L \bigg( -\frac{1}{2}\ga \!\cdot\! \hat F + iD \bigg)\lm, \\
D_-^2 &= D^2 - \hat F^- \!\cdot\! \hat F^- - 2\bar\lm P_L\Slash{D}\lm.
\end{align}
Note that $\Delta=1$ under the Einstein frame gauge $s_0=\bar s_0=e^{-K/6}$.
The symbol $\bbX_{\rm 2f}$ is minus the two-fermion part of the lowest component of $T(\bar w'^2)$,
\begin{align}
\bbX_{\rm 2f} &= - (\hat F^-\!\cdot\!\hat F^-)_{\rm 2f} - 2\bar\lm P_L(\Slash\cD\lm)_{\rm 1f}
+ {2}\bigg( \frac{F_0}{s_0}-\frac{K_\phi F}{3} \bigg) \bar\lm P_L \lm
+ \frac{2}{3}K_\phi \bar\chi P_L\Lm_{\rm 1f}.
\end{align}
Combining these results, we find that the components of $\cR$ we need are given by
\begin{align}
(\cR)_{B\mu} &= \De\frac{i}{(D_-^2)_{\rm 0f}(D_+^2)_{\rm 0f}}
\bar\Lm_{\rm 1f} P_L\ga_\mu \Lm_{\rm 1f},
\end{align}
\begin{align}
(\cR)_{\lm} &= \De \cdot {\sqrt{2}i}
\bigg[ \frac{P_R\Lm_{\rm 1f}}{(D_+^2)_{\rm 0f}} - \frac{P_L\Lm_{\rm 1f}}{(D_-^2)_{\rm 0f}} \bigg],
\end{align}
\begin{align}
(\cR)_D = &
\frac{2\De}{(D_-^2)_{\rm 0f}} \bigg[
 \left( \frac{3F_0}{s_0}-K_\phi F-\frac{\De^2F_T}{(D_+^2)_{\rm 0f}} \right) 
\bar\lm P_L\lm
+ \bar\Lm_{\rm 1f} P_L
\left( K_\phi \chi + \frac{\De^{2}}{(D_+^2)_{\rm 0f}} \chi_T \right)
\bigg] + {\rm h.c.}
 \nn\\
&{}
+2{\De}\bigg(
\frac{(D_-^2)_{\rm 2f}-\bbX_{\rm 2f}}{(D_-^2)_{\rm 0f}}
+ \frac{(D_+^2)_{\rm 2f}-\bar\bbX_{\rm 2f}}{(D_+^2)_{\rm 0f}}
- \frac{\bar\Lm_{\rm 1f} (\Slash\cD\Lm)_{\rm 1f}}{2(D_{-}^2)_{\rm 0f} (D_{+}^2)_{\rm 0f}}
\bigg),
\end{align}
where we only kept two-fermion contributions in $(\cR)_{B\mu}, (\cR)_D$ and one-fermion contributions in $(\cR)_\lm$.
Note also that $(D_-^2)_{\rm 0f}=D^2 - F^- \!\cdot\! F^-$.

\end{document}